\RequirePackage[mathlines]{lineno}
\documentclass[prl,twocolumn,showpacs,amsmath,amssymb]{revtex4-1}
\usepackage{overpic,graphicx}
\usepackage{dcolumn}
\usepackage{bm}
\usepackage{rotating}
\usepackage{subfigure}
\usepackage{color}

\begin{document}

\title{\bf \boldmath
Determination of the pseudoscalar decay constant $f_{D_s^+}$ via $D_s^+\to\mu^+\nu_\mu$
}

\author{M.~Ablikim$^{1}$, M.~N.~Achasov$^{9,d}$, S. ~Ahmed$^{14}$, M.~Albrecht$^{4}$, M.~Alekseev$^{56A,56C}$, A.~Amoroso$^{56A,56C}$, F.~F.~An$^{1}$, Q.~An$^{53,43}$, Y.~Bai$^{42}$, O.~Bakina$^{26}$
% ...
\\
\vspace{0.2cm}
(BESIII Collaboration)\\
\vspace{0.2cm} {\it
$^{1}$ Institute of High Energy Physics, Beijing 100049, People's Republic of China\\
% ...
$^{4}$ Bochum Ruhr-University, D-44780 Bochum, Germany\\
% ...
$^{9}$ G.I. Budker Institute of Nuclear Physics SB RAS (BINP), Novosibirsk 630090, Russia\\
% ...
$^{14}$ Helmholtz Institute Mainz, Johann-Joachim-Becher-Weg 45, D-55099 Mainz, Germany\\
% ...
$^{26}$ Joint Institute for Nuclear Research, 141980 Dubna, Moscow region, Russia\\
% ...
$^{42}$ Southeast University, Nanjing 211100, People's Republic of China\\
$^{43}$ State Key Laboratory of Particle Detection and Electronics, Beijing 100049, Hefei 230026, People's Republic of China\\
% ...
$^{53}$ University of Science and Technology of China, Hefei 230026, People's Republic of China\\
% ...
$^{56}$ (A)University of Turin, I-10125, Turin, Italy; (B)University of Eastern Piedmont, I-15121, Alessandria, Italy; (C)INFN, I-10125, Turin, Italy\\
% ...
$^{d}$ Also at the Novosibirsk State University, Novosibirsk, 630090, Russia\\
% ...
}
}

\begin{abstract}
Using a $3.19~\mathrm{fb}^{-1}$ data sample collected at an $e^+e^-$ center-of-mass energy of $E_{\rm cm}=4.178$\,GeV with the BESIII detector,
we measure 
the branching fraction of the leptonic decay $D_s^+\to\mu^+\nu_\mu$ 
to be
$\mathcal{B}_{D_s^+\to\mu^+\nu_\mu}=(5.49\pm0.16_{\rm stat.}\pm0.15_{\rm syst.})\times10^{-3}$.
Combining our branching fraction with the masses of the $D_s^+$ and $\mu^+$ and the lifetime of the $D_s^+$, 
we determine $f_{D_s^+}|V_{cs}|=246.2\pm3.6_{\rm stat.}\pm3.5_{\rm syst.}~\mathrm{MeV}$.
Using the $c\to s$ quark mixing matrix element $|V_{cs}|$ determined from a global
standard model fit, we evaluate the $D_s^+$ decay constant $f_{D_s^+}=252.9\pm3.7_{\rm stat.}\pm3.6_{\rm syst.}$\,MeV.
Alternatively, using the value of $f_{D_s^+}$ calculated by lattice quantum chromodynamics,
we find $|V_{cs}| = 0.985\pm0.014_{\rm stat.}\pm0.014_{\rm syst.}$.
These values of $\mathcal{B}_{D_s^+\to\mu^+\nu_\mu}$, $f_{D_s^+}|V_{cs}|$, $f_{D_s^+}$ and $|V_{cs}|$ are each the most precise results to date.
\end{abstract}


\maketitle

\oddsidemargin  -0.2cm
\evensidemargin -0.2cm

The leptonic decay $D^+_s\to \ell^+\nu_\ell$~($\ell=e$, $\mu$ or $\tau$)
offers a unique window into both strong and weak effects in the charm quark sector.
In the standard model~(SM), the partial width of the decay $D^+_s\to \ell^+\nu_\ell$
can be written as~\cite{decayrate}
\begin{equation}
\Gamma_{D^+_{s}\to\ell^+\nu_\ell}=\frac{G_F^2}{8\pi}|V_{cs}|^2
f^2_{D^+_{s}}
m_\ell^2 m_{D^+_{s}} \left (1-\frac{m_\ell^2}{m_{D^+_{s}}^2} \right )^2,
\end{equation}
where
$f_{D^+_{s}}$ is the $D^+_{s}$ decay constant,
$|V_{cs}|$ is the $c\to s$
Cabibbo-Kobayashi-Maskawa~(CKM) matrix element,
$G_F$ is the Fermi coupling constant,
$m_\ell$ is the lepton mass, and
$m_{D^+_{s}}$ is the $D^+_{s}$ mass.
In recent years, much progress has been achieved in the measurements of
$f_{D^+_{s}}$ and $|V_{cs}|$
with $D^+_s\to \ell^+\nu_\ell$ decays at the
CLEO~\cite{cleo2009,cleo2009a,cleo2009b}, BaBar~\cite{babar2010}, Belle~\cite{belle2013} and
BESIII~\cite{bes2016} experiments.
However, compared to the precision of the most accurate 
lattice quantum chromodynamics~(LQCD)
calculation of $f_{D^+_s}$~\cite{FLab2018}, the accuracy of the  measurements is still limited. 
Improved measurements of $f_{D^+_{s}}$ and $|V_{cs}|$
are critical to calibrate various theoretical calculations of $f_{D^+_{s}}$~\cite{FLab2018,LQCD,etm2015,ukqcd2017,ukqcd2015,milc2012,hpqcd2010,hpqcd2012,milc2005,hpqcd2008,etm2012,chen2014,pacs2011,qcdsf2007,chiu2005,ukqcd2001,becirevic1999,bordes2005,narison2002,badalian2007,ebert2006,cvetic2004,choi2007,salcedo2004,wang2004,amundson1993,becirevic2013,lucha2011,hwang2010,wang2015},
such as those from quenched and unquenched LQCD, QCD sum rules, $etc.$,
and to test the unitarity of the quark mixing matrix with better precision.

In the SM, the ratio of the branching fraction (BF) of $D^+_s\to \tau^+\nu_\tau$
over that of $D^+_s\to \mu^+\nu_\mu$ is predicted to be 9.74 with negligible uncertainty
and the BFs of $D_s^+\to\mu^+\nu_\mu$ and $D_s^-\to\mu^-\bar{\nu}_\mu$ decays
are expected to be the same.
However, hints of lepton flavor universality~(LFU) violation in semileptonic $B$ decays were recently reported at BaBar, LHCb and Belle~\cite{babar_1,babar_2,lhcb_1,lhcb_kee_3,belle_kee}.
It has been argued that new physics mechanisms, such as a two-Higgs-doublet model with the mediation of charged Higgs bosons~\cite{fajfer,2hdm} or a Seesaw mechanism due to lepton mixing with Majorana neutrinos~\cite{seesaw}, may cause LFU or CP violation.
Tests of LFU and searches for CP violation 
in $D^+_s\to\ell^+\nu_\ell$ decays are therefore important tests of the SM.

In this Letter, we present an experimental study of the leptonic decay $D_s^+\to\mu^+\nu_\mu$~\cite{conjugate}
by analyzing a 3.19\,fb$^{-1}$ data sample
collected with the BESIII detector 
at an $e^+e^-$ center-of-mass energy of $E_{\rm cm}=4.178$\,GeV. At this energy,
$D^+_s$ mesons are produced mainly through the process $e^+e^-\to D^+_sD_s^{*-}+c.c$.
In an event where a $D_s^-$ meson (called a single-tag~(ST) $D_s^-$ meson)
is fully reconstructed, one can then search for a $\gamma$ or $\pi^0$ and a 
$D_s^+$ meson in the recoiling system (called a double-tag (DT) event).

Details about the design and performance of the BESIII detector are given in Ref.~\cite{BESCol}.
The endcap time-of-flight~(TOF) system was upgraded with multi-gap resistive plate chamber technology and now has a time resolution of 60~ps~\cite{mrpc1,mrpc2}.
Monte Carlo~(MC) events are generated with a {\sc geant}4-based~\cite{geant4} detector simulation software package~\cite{boost}, which includes both the geometrical description of the detector and the detector's response. An inclusive MC sample is produced at $E_{\rm cm}=4.178$\,GeV, which includes all open charm processes,
initial state radiation~(ISR) production of the $\psi(3770)$, $\psi(3686)$ and $J/\psi$,
and $q\bar{q}\,(q=u,d,s)$ continuum processes, along with Bhabha scattering,
$\mu^+\mu^-$, $\tau^+\tau^-$ and $\gamma\gamma$ events.
The open charm processes are generated using {\sc ConExc}~\cite{conexc}. 
The effects of ISR~\cite{isr} and
final state radiation~(FSR)~\cite{photons} are considered.
The 
decay modes with known BF are generated using {\sc EvtGen}~\cite{evtgen} and the other modes are
generated using {\sc LundCharm}~\cite{lundcharm}.

The ST $D^-_s$ mesons are reconstructed from 14 hadronic decay modes, 
$D^-_s\to K^+K^-\pi^-$, $K^+K^-\pi^-\pi^0$, $K^0_SK^-$,
$K^0_SK^-\pi^0$,
$K^0_SK^0_S\pi^-$,
$K^0_SK^+\pi^-\pi^-$,
$K^0_SK^-\pi^+\pi^-$,
$K^-\pi^+\pi^-$,
$\pi^+\pi^-\pi^-$,
$\eta_{\gamma\gamma}\pi^-$, $\eta_{\pi^0\pi^+\pi^-}\pi^-$,
$\eta^\prime_{\eta_{\gamma\gamma}\pi^+\pi^-}\pi^-$, $\eta^\prime_{\gamma\rho^0}\pi^-$, and
$\eta_{\gamma\gamma}\rho^-$,
where the subscripts of $\eta^{(\prime)}$ represent the decay modes used to reconstruct $\eta^{(\prime)}$.

All charged tracks except for those from $K_S^0$ decays must originate from the interaction
point~(IP) with a distance of closest approach less than 1 cm in the transverse plane 
and less than 10 cm along the $z$ axis. The polar angle $\theta$ of each track defined with respect to the positron beam
must satisfy $|\cos\theta|<0.93$. Measurements of the specific ionization energy
loss~($dE/dx$) in the main drift chamber and the TOF
are combined and used for particle identification~(PID) by forming
confidence levels for pion and kaon hypotheses ($CL_\pi$,~$CL_K$).
Kaon~(pion) candidates are required to satisfy $CL_{K(\pi)}>CL_{\pi(K)}$.

To select $K_S^0$ candidates, pairs of oppositely charged tracks with distances of closest approach to the IP less than 20 cm along the $z$ axis are assigned as
$\pi^+\pi^-$ without PID requirements.
These $\pi^+\pi^-$ combinations are required to have an invariant mass within $\pm12$\,MeV of the nominal $K_S^0$ mass~\cite{PDG2017} and
have a decay length of the reconstructed $K_S^0$ larger than $2\sigma$ of the vertex resolution away from the IP.
The $\pi^0$ and $\eta$ mesons are reconstructed via $\gamma\gamma$ decays.
It is required that each electromagnetic shower starts within 700\,ns of the event start time and
its energy is greater than 25\,(50)\,MeV
in the barrel\,(endcap) region of the electromagnetic calorimeter~(EMC)~\cite{BESCol}.
The opening angle between the shower and
the nearest charged track has to be greater than $10^{\circ}$.
The $\gamma\gamma$ combinations with an invariant mass $M_{\gamma\gamma}\in(0.115,\,0.150)$
and $(0.50,\,0.57)$\,GeV$/c^{2}$ are regarded as $\pi^0$ and $\eta$ mesons, respectively.
A kinematic fit is performed to constrain $M_{\gamma\gamma}$ to the $\pi^{0}$ or $\eta$
nominal mass~\cite{PDG2017}.
The $\eta$ candidates for the $\eta\pi^-$ 
ST channel are also reconstructed
via $\pi^0\pi^+\pi^-$ candidates with an invariant mass
within $(0.53,\,0.57)~\mathrm{GeV}/c^2$.
The $\eta^\prime$ mesons are reconstructed via two decay modes, $\eta\pi^+\pi^-$ and $\gamma\rho^0$,
whose invariant masses are required to be within
$(0.946,\,0.970)$ and $(0.940,\,0.976)~\mathrm{GeV}/c^2$, respectively.
In addition, the minimum energy
of the $\gamma$ from $\eta'\to\gamma\rho^0$ decays must be greater than 0.1\,GeV.
The $\rho^0$ and $\rho^+$ mesons are reconstructed from $\pi^+\pi^-$ and $\pi^+\pi^0$
candidates, whose invariant masses are required to be larger than $0.5~\mathrm{GeV}/c^2$
and within $(0.67,\,0.87)~\mathrm{GeV}/c^2$, respectively.

The momentum of any pion not originating from a $K_S^0$, $\eta$, or $\eta^\prime$ decay is required to be greater than 0.1\,GeV/$c$
to reject soft pions from $D^*$ decays.
For $\pi^+\pi^-\pi^-$ and $K^-\pi^+\pi^-$ combinations, the dominant peaking backgrounds from $K^0_S\pi^-$ and $K_S^0K^-$ events are rejected by requiring the invariant mass of any $\pi^+\pi^-$ combination be more than $\pm 0.03$ GeV/$c^2$ away from the nominal $K^0_S$ mass~\cite{PDG2017}.

\begin{figure}[htbp]
\centering
\includegraphics[width=0.48\textwidth]
{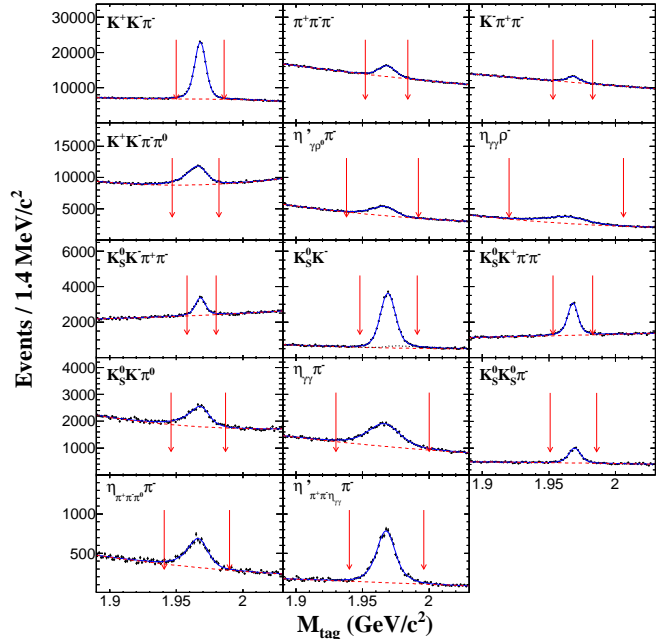}
\caption{\footnotesize
Fits to the $M_{\rm tag}$ distributions of the accepted ST candidates.
Dots with error bars are data.
Blue solid curves are the fit results.
Red dashed curves are the fitted backgrounds.
The black dotted curve in the $K_S^0K^-$ mode is the $D^-\to K_S^0\pi^-$ component.
The pairs of arrows denote the signal regions.
}
\label{fig:stfit}
\end{figure}

To suppress non-$D_s^+D^{*-}_s$ events,
the beam-constrained mass of the ST $D_s^-$
candidate 
\begin{equation}
M_{\rm BC}\equiv\sqrt{(E_{\rm cm}/2)^2-|\vec{p}_{D_s^-}|^2}
\end{equation}
is required to be within $(2.010,\,2.073)\,\mathrm{GeV}/c^2$,
where
$\vec{p}_{D_s^-}$ is the momentum of the ST $D_s^-$ candidate.
This requirement retains $D_s^-$ mesons directly from $e^+e^-$ annihilation and indirectly from $D_s^{*-}$
decay (See Fig. 1 in Ref.~\cite{supplemental}).
In each event, we only keep
the candidate with the $D_s^-$ recoil mass 
\begin{equation}
M_{\rm rec} \equiv \sqrt{ \left (E_{\rm cm} - \sqrt{|\vec p_{D^-_s}|^2+m^2_{D^-_s}} \right )^2
-|\vec p_{D^-_s}|^2}
\end{equation}
closest to the nominal $D_s^{*+}$ mass~\cite{PDG2017} per tag mode per charge.
Figure~\ref{fig:stfit} shows the invariant mass ($M_{\rm tag}$) spectra of the accepted ST
candidates.
The ST yield for each tag mode is obtained by a fit to
the corresponding $M_{\rm tag}$ spectrum.
The signal is described by the MC-simulated shape convolved with a Gaussian function
representing the resolution difference between data and MC simulation.
For the tag mode $D^-_s\to K_S^0K^-$,
the peaking background from $D^-\to K^0_S\pi^-$ is described by the MC-simulated shape and then smeared with the same Gaussian function used in the signal shape with its size as a free parameter.
The non-peaking background is modeled by a second- or third-order Chebychev polynomial
function. Studies of the inclusive MC sample validate this parametrisation of the background shape.
The fit results on these invariant mass spectra are shown in Fig.~\ref{fig:stfit}.
The events in the signal regions 
are kept for further analysis.
The total ST yield in data is $N^{\rm tot}_{\rm ST}=388660\pm2592$ (see tag-dependent
ST yields and background yields in the signal regions in Table~I of Ref.~\cite{supplemental}).

\begin{figure}[htbp]
\centering
\includegraphics[height=5cm]
{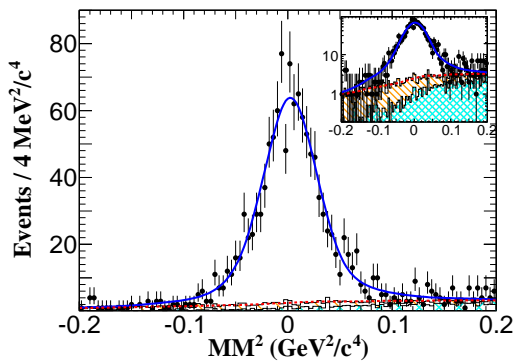}
\caption{\footnotesize
Fit to the $\mathrm{MM}^2$ distribution of the $D^+_s\to \mu^+\nu_\mu$ candidates.
Inset plot shows the same distribution in log scale.
Dots with error bars are data.
Blue solid curve is the fit result.
Red dotted curve is the fitted background.
Orange hatched and blue cross-hatched histograms are the BKGI component and the combined BKGII and BKGIII components, respectively (see text).
}
\label{fig:mm2fit}
\end{figure}

At the recoil sides of the ST $D_s^-$ mesons,
the $D_s^+\to\mu^+\nu_\mu$ candidates are selected with the surviving neutral and charged tracks.
To select the soft $\gamma(\pi^0)$ from $D_s^{*}$ and
to separate signals from combinatorial backgrounds,
we define two kinematic variables 
\begin{equation}
\Delta E \equiv E_{\rm cm}-E_{\rm tag}-E_{\rm miss}-E_{\gamma(\pi^0)} 
\end{equation}
and
\begin{align}
\mathrm{MM}^2&\equiv\left (E_{\rm cm}-E_{\rm tag}-E_{\gamma(\pi^0)}-E_{\mu}\right )^2\nonumber\\
&-|-\vec{p}_{\rm tag}-\vec{p}_{\gamma(\pi^0)}-\vec{p}_{\mu}|^2.
\end{align}
Here
$E_{\rm miss} \equiv \sqrt{|\vec{p}_{\rm miss}|^2+m_{D_s^+}^2}$ and
$\vec{p}_{\rm miss} \equiv -\vec{p}_{\rm tag}-\vec{p}_{\gamma(\pi^0)}$ are
the missing energy and momentum of the recoiling system of the soft $\gamma(\pi^0)$ and the ST $D_s^-$,
where $E_i$ and $\vec p_i$ ($i=\mu,\gamma(\pi^0)$ or tag) denote
the energy and momentum of the muon, $\gamma(\pi^0)$ or ST $D^-_s$, respectively.
$\mathrm{MM}^2$ is the missing mass square of the undetectable neutrino.
We loop over all remaining $\gamma$ or $\pi^0$ candidates 
and choose the one giving a minimum $|\Delta E|$.
The events with $\Delta E\in(-0.05,\,0.10)$\,GeV are accepted. 
The muon candidate is required to have an opposite charge 
to the ST $D^-_s$ meson and a
deposited energy in the EMC within $(0.0,\,0.3)$\,GeV.
It must also satisfy a two dimensional (2D, e.g., $|\cos\theta_\mu|$ and momentum $p_{\mu}$) requirement on the hit depth ($d_\mu$) in the muon counter, as explained in Ref.~\cite{muid}.
To suppress the backgrounds with extra photon(s),
the maximum energy of the unused showers in the DT
selection ($E_{\mathrm{extra}~\gamma}^{\rm max}$) is required to be less than 0.4\,GeV
and no additional charged track that satisfies the charged track selection criteria is allowed.
To improve the $\mathrm{MM}^2$ resolution,
the candidate tracks, plus the missing neutrino, are subjected to a 4-constraint kinematic fit requiring energy and momentum conservation.
In addition, the invariant masses of the two $D_s$ mesons are constrained to the nominal $D_s$ mass,
the invariant mass of the $D_s^-\gamma(\pi^0)$ or $D_s^+\gamma(\pi^0)$ combination is
constrained to the nominal $D_s^*$ mass,
and the combination with the smaller $\chi^2$ is kept.
Figure~\ref{fig:mm2fit} shows the $\mathrm{MM}^2$ distribution for the accepted
DT candidate events.

To extract the DT yield, an unbinned constrained fit is performed to the $\mathrm{MM}^2$ distribution.
In the fit, the background events are classified into three categories:
events with correctly reconstructed ST $D_s^-$ and $\mu^+$ but an unmatched $\gamma(\pi^0)$ from the $D_s^{*-}$~(BKGI),
events with a correctly reconstructed ST $D_s^-$ but misidentified $\mu^+$~(BKGII), and
other events with a misreconstructed ST $D_s^-$~(BKGIII).
The signal and BKGI shapes are modeled with MC simulation.
The signal shape is convolved with a Gaussian function with its mean and width as free parameters.
The ratio of the signal yield over the BKGI yield
is constrained to the value determined with the signal MC events.
The size and shape of the BKGII and BKGIII components
are fixed by analyzing the inclusive MC sample.
From the fit to the $\mathrm{MM}^2$ distribution, as shown in Fig.~\ref{fig:mm2fit},
we determine the number of $D_s^+\to\mu^+\nu_\mu$ decays
to be $N_{\rm DT}=1135.9\pm33.1$.

The efficiencies for reconstructing the DT candidate events
are determined with an exclusive MC sample of $e^+e^-\to D_s^+D^{*-}_s$, where
the $D_s^-$ decays to each tag mode and the $D_s^+$ decays to $\mu^+\nu_\mu$.
Dividing them by the ST efficiencies determined with the inclusive MC sample yields the corresponding
efficiencies of the $\gamma(\pi^0)\mu^+\nu_\mu$ reconstruction. 
The averaged efficiency of finding $\gamma(\pi^0)\mu^+\nu_\mu$ is 
$(52.67\pm0.19)\%$ as determined from
\begin{equation}
\varepsilon_{\gamma(\pi^0)\mu^+\nu_\mu}=f_{\mu\,\rm PID}^{\rm cor}\sum_{i}(N_{\rm ST}^i\varepsilon_{\rm DT}^i)/(N_{\rm ST}^{\rm tot}\varepsilon_{\rm ST}^i),
\end{equation}
where $N_{\rm ST}^i$, $\varepsilon_{\rm ST}^i$, and $\varepsilon_{\rm DT}^i$ are the ST yield, ST efficiency and DT efficiency in the $i$-th ST mode, respectively. 
The factor $f_{\mu\,\rm PID}^{\rm cor}=0.897$ accounts for 
the difference between the $\mu^+$ PID efficiencies in data and MC simulation [$
\varepsilon_{\mu\,\rm PID}^{\rm data\,(MC)}$].
These efficiencies are estimated using $e^+e^-\to\gamma\mu^+\mu^-$ samples but reweighted by
the $\mu^+$ 2D distribution of $D_s^+\to\mu^+\nu_\mu$. It is nonnegligible 
mainly due to the imperfect simulation of $d_\mu$ and its applicability in 
different topology environments is verified via
three aspects: 
(1) Studies with signal MC events show that $\varepsilon_{\mu\,\rm PID}^{\rm MC}=(74.79\pm0.03)\%$ for 
$D_s^+\to\mu^+\nu_\mu$ signals can be well reproduced by the
2D reweighted efficiency $\varepsilon_{\mu\,\rm PID}^{\rm MC}=(74.91\pm0.10)\%$ with $e^+e^-\to\gamma\mu^+\mu^-$ samples.
(2) Our nominal BF ($\mathcal{B}_{D_s^+\to\mu^+\nu_\mu}$) obtained later can be well reproduced by removing the $d_\mu$ 
requirement, with negligible difference but obviously lower precision due to much higher background~\cite{dstauv}.
(3) The $\varepsilon_{\mu\,\rm PID}^{\rm data\,(MC)}$ for 
$e^+e^-\to\gamma_{\rm ISR}\psi(3686),\psi(3686)\to\pi^+\pi^-J/\psi,J/\psi\to\mu^+\mu^-$ events
can be well reproduced by the corresponding
2D reweighted efficiencies with $e^+e^-\to\gamma\mu^+\mu^-$ samples
(see Table II of Ref.~\cite{supplemental}).
The BF of $D_s^+\to\mu^+\nu_\mu$ is then determined 
to be $(5.49\pm0.16_{\rm stat.}\pm0.15_{\rm syst.})\times10^{-3}$ from
\begin{equation}
\mathcal{B}_{D_s^+\to\mu^+\nu_\mu}=f_{\rm cor}^{\rm rad}N_{\rm DT}/(N_{\rm ST}^{\rm tot}\varepsilon_{\gamma(\pi^0)\mu^+\nu_\mu}),
\end{equation}
where the radiative correction factor $f_{\rm cor}^{\rm rad}=0.99$ is due to the contribution from 
$D^+_s\to \gamma {\mathcal D}^{*+}_s \to \gamma \mu^+\nu_\mu$~\cite{radiation},
with ${\mathcal D}^{*+}_s$ as a virtual vector or axial-vector meson.
This contribution is almost identical with our signal process for low energy radiated photons.
We further examine the BFs measured with individual tags which have very different 
background levels, and a good consistence is found (see Table I of Ref.~\cite{supplemental} for
tag-dependent DT yields, $\varepsilon_{\gamma(\pi^0)\mu^+\nu_\mu}$ and $\mathcal{B}_{D_s^+\to\mu^+\nu_\mu}$).

The systematic uncertainties in the BF measurement are estimated relative to the measured BF and are described below.

For uncertainties in the event selection criteria,
the $\mu^+$ tracking and PID efficiencies are studied with $e^+e^-\to\gamma\mu^+\mu^-$ events. 
After correcting the detection efficiency by $f^{\rm cor}_{\mu\,\rm PID}$, we assign 0.5\% and 0.8\%
as the uncertainties in $\mu^+$ tracking and PID efficiencies, respectively.
The photon reconstruction efficiency has been previously studied with $J/\psi\to\pi^+\pi^-\pi^0$ decays~\cite{geff}. The uncertainty of finding $\gamma(\pi^0)$ is weighted according to the BFs of $D_s^{*+}\to\gamma D_s^+$ and $D_s^{*+}\to\pi^0D_s^+$~\cite{PDG2017} and assigned to be 1.0\%.
The efficiencies for the requirements of $E_{\mathrm{extra}~\gamma}^{\rm max}$ and
no extra good charged track 
are studied with a DT hadronic sample.
The systematic uncertainties are taken to be 0.3\% and 0.9\% considering the efficiency differences between data and MC simulation, respectively.
The uncertainty of the $\Delta E$ requirement is estimated by varying the signal region by
$\pm0.01$\,GeV, and the maximum
change of the BF, 0.5\%, is taken as the systematic uncertainty.

To determine the uncertainty in the $\mathrm{MM}^2$ fit, we change the fit range
by $\pm0.02~\mathrm{GeV}^2/c^4$, and the largest change
of the BF is 0.6\%. We change the signal shape by varying the $\gamma(\pi^0)$
match requirement and the maximum change is 0.2\%. 
Two sources of uncertainty in the background estimation are considered.
The effect of the background shape is obtained to be 0.2\% by shifting the number
of the main components of BKGII by $\pm 1\sigma$ of the uncertainties
of the corresponding BFs~\cite{PDG2017},
and varying the relative fraction of the main components of BKGII by 50\%.
The effect of the fixed number of the BKGII and BKGIII is estimated to be 0.5\% by varying
the nominal numbers by $\pm 1\sigma$ of their uncertainties.
To evaluate the uncertainty in the fixed ratio of signal and BKGI,
we perform an alternative fit to the $\mathrm{MM}^2$ distribution of data
without constraining the ratio of signal and BKGI.
The change in the DT yield, 1.1\%, is assigned as the relevant uncertainty.

The uncertainty in the number of ST $D_s^-$ mesons is assigned to be 0.8\% by examining the changes
of the fit yields when varying the signal shape, background shape, bin size and fit range and considering the background fluctuation in the fit.
The uncertainty due to the limited MC size is 0.4\%.
The uncertainty in the imperfect simulation of the FSR effect is estimated as 0.4\% by varying the amount of FSR photons in signal MC events~\cite{photons}.
The uncertainty due to the quoted BFs of $D_s^{*-}$ subdecays from the particle data group~(PDG)~\cite{PDG2017} is examined by varying each subdecay BF by $\pm 1\sigma$. The efficiency change is found to be 0.4\% and is taken as the associated uncertainty.
The uncertainty in the radiative correction is assigned to be 1.0\%,
which is taken as 100\% of its central value from theoretical calculation~\cite{radiation}.
The ST efficiencies in the inclusive and signal MC samples are slightly different with each other due to different track multiplicities in these two environments. This may cause incomplete cancellation of the uncertainties of the ST efficiencies. The associated uncertainty is assigned as 0.6\%, by taking into account the differences of the efficiencies of tracking/PID of $K^\pm$ and $\pi^\pm$, as well as the selections of neutral particles between data and MC simulation in different environments.
The total systematic uncertainty is determined to be 2.7\% by adding all the uncertainties in quadrature.

Combining our BF with
the world average values of $G_F$, $m_\mu$, $m_{D^+_s}$ and
the lifetime of $D_s^+$~\cite{PDG2017} in Eq. (1) yields
$$f_{D_s^+}|V_{cs}|=246.2\pm3.6_{\rm stat.}\pm3.5_{\rm syst.}~\mathrm{MeV}.$$
Here the systematic uncertainties arise mainly from the uncertainties in the measured
BF (1.5\%) and the lifetime of the $D^+_s$ (0.4\%).
Taking the CKM matrix element $|V_{cs}|=0.97359_{-0.00011}^{+0.00010}$ from the global fit
in the SM~\cite{PDG2017} or the averaged decay constant
$f_{D_s^+}=249.9\pm0.4~\mathrm{MeV}$ of recent LQCD calculations~\cite{FLab2018,etm2015}
as input, we determine
$$f_{D_s^+}=252.9\pm3.7_{\rm stat.}\pm3.6_{\rm syst.}~\mathrm{MeV}$$
and
$$|V_{cs}|=0.985\pm0.014_{\rm stat.}\pm0.014_{\rm syst.}.$$
The additional systematic uncertainties according to the input parameters are
negligible for $|V_{cs}|$ and 0.2\% for $f_{D_s^+}$.
The measured $|V_{cs}|$ is consistent with our measurements using 
$D\to\bar K\ell^+\nu_\ell$~\cite{bes3_kev,bes3_ksev,bes3_klev,bes3_kmuv} and $D_s^+\to\eta^{(\prime)}e^+\nu_e$~\cite{bes3_etaev}, but with
much better precision.

Combining the obtained $f_{D_s^+}|V_{cs}|$ and its counterpart $f_{D^+}|V_{cd}|$
measured in our previous work~\cite{fdp},
along with $|V_{cd}/V_{cs}|=0.23047\pm0.00045$ from the SM global fit~\cite{PDG2017},
yields $f_{D_s^+}/f_{D^+}=1.24\pm0.04_{\rm stat.}\pm0.02_{\rm syst.}$. It is consistent with the CLEO measurement~\cite{cleo2009} within 1$\sigma$ and 
the LQCD calculation within 2$\sigma$~\cite{FLab2018}.
Alternatively, with the input of $f_{D_s^+}/f_{D^+}=1.1749\pm0.0016$ calculated by LQCD~\cite{FLab2018},
we obtain $|V_{cd}/V_{cs}|^2=0.048\pm0.003_{\rm stat.}\pm0.001_{\rm syst.}$, which agrees with
the one expected by $|V_{cs}|$ and $|V_{cd}|$ given by the CKMfitter within 2$\sigma$.
Here, only the systematic uncertainty in the radiative correction is canceled since the two data samples
were taken in different years.

Based on our result for ${\mathcal{B}}_{D_s^+\to\mu^+\nu_\mu}$ and those measured at the
CLEO~\cite{cleo2009}, BaBar~\cite{babar2010} and Belle~\cite{belle2013} experiments, 
along with a previous measurement at BESIII~\cite{bes2016},
the inverse-uncertainty weighted BF is determined to be
$\bar {\mathcal{B}}_{D_s^+\to\mu^+\nu_\mu}=(5.49\pm0.17)\times10^{-3}$~\cite{bfweight}.
The ratio of $\bar {\mathcal{B}}_{D_s^+\to\mu^+\nu_\mu}$
over the PDG value of
$\mathcal{B}_{D_s^+\to\tau^+\nu_\tau}=(5.48\pm0.23)\%$~\cite{PDG2017} is determined to be
$\frac{\mathcal{B}_{D_s^+\to\tau^+\nu_\tau}}{\bar {\mathcal{B}}_{D_s^+\to\mu^+\nu_\mu}}=9.98\pm0.52,$
which agrees with the SM predicted value of 9.74 within uncertainty.

The BFs of $D_s^+\to\mu^+\nu_\mu$ and $D_s^-\to\mu^-\bar{\nu}_\mu$ decays are also measured separately.
The results are ${\mathcal B}_{D_s^+\to\mu^+\nu_\mu}=(5.62\pm0.23_{\rm stat.})\times10^{-3}$ and ${\mathcal B}_{D_s^-\to\mu^-\bar \nu_\mu}=(5.40\pm0.23_{\rm stat.})\times10^{-3}$.
The BF asymmetry is determined to be
$A_{\rm CP}=\frac{{\mathcal B}_{D_s^+\to\mu^+\nu_\mu}-{\mathcal B}_{D_s^-\to\mu^-\bar{\nu}_\mu}}{{\mathcal B}_{D_s^+\to\mu^+\nu_\mu}+{\mathcal B}_{D_s^-\to\mu^-\bar{\nu}_\mu}}=(2.0\pm3.0_{\rm stat.}\pm1.2_{\rm syst.})\%,$
where the uncertainties in the tracking and PID efficiencies of the muon,
the ST yields, the limited MC statistics, as well as the signal shape and fit range in $\mathrm{MM}^2$ fits for $D_s^+$ and $D_s^-$ have been studied separately and are not canceled.

In summary, by analyzing 3.19~fb$^{-1}$ of $e^+e^-$ collision data 
collected at $E_{\rm cm}=4.178$\,GeV with the BESIII detector,
we have measured $\mathcal{B}(D^+_s\to\mu^+\nu_\mu)$, the decay constant $f_{D_s^+}$, and the CKM matrix element $|V_{cs}|$. These are the most precise measurements to date,
and are important to calibrate various theoretical calculations of $f_{D_s^+}$ and test the unitarity of the CKM matrix with better accuracy.
We also search for LFU and CP violation in $D_s^+\to\ell^+\nu_\ell$ decays, and no evidence is found.

The BESIII collaboration thanks the staff of BEPCII and the IHEP computing center for their strong support. This work is supported in part by National Key Basic Research Program of China under Contract No. 2015CB856700; National Natural Science Foundation of China (NSFC) under Contracts Nos. 11235011, 11335008, 11425524, 11505034, 11575077, 11625523, 11635010, 11675200, 11775230; the Chinese Academy of Sciences (CAS) Large-Scale Scientific Facility Program; the CAS Center for Excellence in Particle Physics (CCEPP); Joint Large-Scale Scientific Facility Funds of the NSFC and CAS under Contracts Nos. U1332201, U1532257, U1532258, U1632109; CAS under Contracts Nos. KJCX2-YW-N29, KJCX2-YW-N45, QYZDJ-SSW-SLH003; 100 Talents Program of CAS; National 1000 Talents Program of China; INPAC and Shanghai Key Laboratory for Particle Physics and Cosmology; German Research Foundation DFG under Contracts Nos. Collaborative Research Center CRC 1044, FOR 2359; Istituto Nazionale di Fisica Nucleare, Italy; Koninklijke Nederlandse Akademie van Wetenschappen (KNAW) under Contract No. 530-4CDP03; Ministry of Development of Turkey under Contract No. DPT2006K-120470; National Science and Technology fund; The Swedish Research Council; U. S. Department of Energy under Contracts Nos. DE-FG02-05ER41374, DE-SC-0010118, DE-SC-0010504, DE-SC-0012069; University of Groningen (RuG) and the Helmholtzzentrum fuer Schwerionenforschung GmbH (GSI), Darmstadt; WCU Program of National Research Foundation of Korea under Contract No. R32-2008-000-10155-0.

\clearpage
\appendix
\onecolumngrid
\section*{Supplemental material}
\setcounter{table}{0}
\setcounter{figure}{0}

\begin{figure*}[htbp]\centering
	\includegraphics[height=7cm]{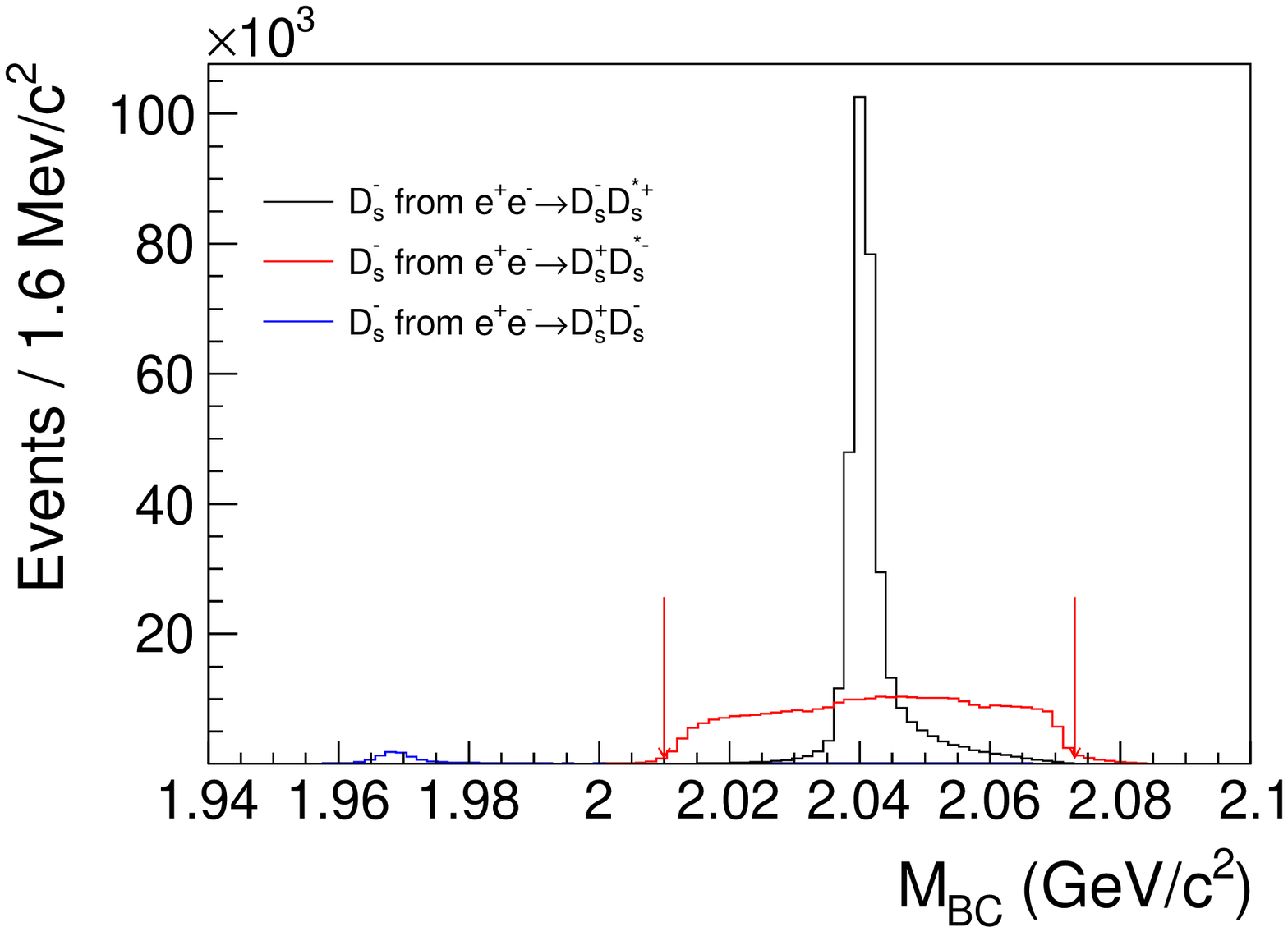}
	\caption{The $M_{\rm BC}$ distributions of the ST $D_s^-$ candidates from $e^+e^-\to D_s^+D_s^{*-}$,
	$e^+e^-\to D_s^-D_s^{*+}$, and $e^+e^-\to D_s^-D_s^+$ processes. The red arrows give our nominal
	$M_{\rm BC}$ window for the ST $D_s^-$ candidates.}
	\label{fig:mBC}
\end{figure*}

Figure~\ref{fig:mBC} shows the $M_{\rm BC}$ distributions of the ST $D_s^-$ candidates from
$e^+e^-\to D_s^-D_s^{*+}$, $e^+e^-\to D_s^+D_s^{*-}$, and $e^+e^-\to D_s^+D_s^-$ processes
based on MC simulation. Both $D_s^-$ mesons directly produced from $e^+e^-$ annihilation and
indirectly produced from $D_s^{*-}$ decays are retained by our nominal $M_{\rm BC}$ requirement.

Table~\ref{tab:bf} summarizes the ST yield $N_{\rm ST}$, the background yield $N_{\rm ST}^{\rm bkg}$
in the $M_{\rm tag}$ signal regions, the DT yield $N_{\rm DT}$, the signal efficiency
$\varepsilon_{\gamma(\pi^0)\mu^+\nu_\mu}$ and the obtained $\mathcal{B}_{D_s^+\to\mu^+\nu_\mu}$
for each ST mode. Although the background levels for various ST modes are much different, the
BFs measured with individual ST modes are consistent with each other.

As an independent check, we further examine the $\mu^+$ PID efficiencies of data and MC simulation,
$\varepsilon_{\mu~\mathrm{PID}}^{\rm data}$ and $\varepsilon_{\mu~\mathrm{PID}}^{\rm MC}$, by
analyzing $e^+e^-\to\gamma_{\rm ISR}\psi(3686),\psi(3686)\to\pi^+\pi^-J/\psi,J/\psi\to\mu^+\mu^-$
events (sample I) and corresponding 2D reweighted efficiencies based on $e^+e^-\to\gamma\mu^+\mu^-$
samples (sample II). Two samples with much different topologies give consistent 
$\varepsilon_{\mu~\mathrm{PID}}^{\rm data}$, $\varepsilon_{\mu~\mathrm{PID}}^{\rm MC}$, and
$f_{\mu~\mathrm{PID}}^{\rm cor}=\varepsilon_{\mu~\mathrm{PID}}^{\rm data}/\varepsilon_{\mu~\mathrm{PID}}^{\rm MC}$,
as shown in Table~\ref{tab:mupid}. The obtained $f_{\mu~\mathrm{PID}}^{\rm cor}$ in these two
samples are different with that in $D_s^+\to\mu^+\nu_\mu$ mainly due to much higher muon momentum.

\begin{table*}[htbp]\centering
	\caption{Summary of $N_{\rm ST}$, $N_{\rm ST}^{\rm bkg}$, $N_{\rm DT}$, $\varepsilon_{\gamma(\pi^0)\mu^+
	\nu_\mu}$, and the obtained $\mathcal{B}_{D_s^+\to\mu^+\nu_\mu}$ with various ST modes. The
	uncertainties are only statistical. The signal efficiencies have been corrected by $f_{\mu~\mathrm{PID}}
	^{\rm cor}$ as described in manuscript. The variations of the signal efficiencies are mainly due
	to different multiplicities of the tag sides.}
	\label{tab:bf}
	\begin{tabular}{lccccc}\hline
		ST mode & $N_{\rm ST}$ & $N_{\rm ST}^{\rm bkg}$ & $N_{\rm DT}$ & $\varepsilon_{\gamma(\pi^0)\mu^+\nu_\mu}$ (\%) & $\mathcal{B}_{D_s^+\to\mu^+\nu_\mu}~(\times10^{-3})$ \\\hline
		$K^+K^-\pi^+$										& $133959\pm633$	& 173160	& $373.3\pm18.9$	& $49.73\pm0.24$ & $5.55\pm0.28$\\
		$K^+K^-\pi^+\pi^0$									& $41377\pm916$		& 221099	& $123.1\pm10.7$ 	& $57.32\pm0.85$ & $5.14\pm0.46$\\
		$\pi^+\pi^+\pi^-$									& $35966\pm913$		& 300499	& $90.0\pm9.9$		& $51.21\pm0.53$ & $4.84\pm0.55$\\
		$K_S^0K^+$											& $32039\pm291$		& 18776		& $79.7\pm9.0$ 		& $49.77\pm0.36$ & $4.95\pm0.56$\\
		$K_S^0K^+\pi^0$										& $11294\pm433$		& 52788		& $38.4\pm6.1$ 		& $56.71\pm2.34$ & $5.94\pm0.97$\\
		$K^+\pi^+\pi^-$										& $15877\pm872$		& 246528	& $45.6\pm7.2$ 		& $51.21\pm1.30$ & $5.55\pm0.93$\\
		$K_S^0K_S^0\pi^+$									& $4832\pm180$		& 11274		& $20.2\pm4.4$ 		& $50.55\pm1.25$ & $8.19\pm1.82$\\
		$K_S^0K^-\pi^+\pi^+$								& $14046\pm240$		& 26873		& $44.1\pm6.5$ 		& $51.91\pm0.91$ & $5.98\pm0.89$\\
		$K_S^0K^+\pi^+\pi^-$								& $7171\pm292$		& 37456		& $24.7\pm4.9$ 		& $54.14\pm1.21$ & $6.29\pm1.28$\\
		$\eta_{\gamma\gamma}\pi^+$							& $19323\pm725$		& 53701		& $63.5\pm8.1$ 		& $52.72\pm0.62$ & $6.17\pm0.82$\\
		$\eta_{\pi^+\pi^-\pi^0}\pi^+$						& $5508\pm202$		& 11225		& $20.2\pm4.5$ 		& $54.00\pm1.13$ & $6.73\pm1.51$\\
		$\eta^\prime_{\pi^+\pi^-\eta_{\gamma\gamma}}\pi^+$	& $9242\pm155$		& 5002		& $33.0\pm5.7$ 		& $56.30\pm0.54$ & $6.27\pm1.09$\\
		$\eta^\prime_{\gamma\rho^0}\pi^+$					& $25191\pm695$		& 152363	& $75.1\pm8.6$ 		& $53.74\pm0.72$ & $5.49\pm0.65$\\
		$\eta_{\gamma\gamma}\rho^+$							& $32835\pm1537$	& 166324	& $108.4\pm10.5$	& $60.70\pm0.91$ & $5.38\pm0.58$\\
		\hline
	\end{tabular}
\end{table*}

\begin{table*}[htbp]\centering
	\caption{Summary of $\varepsilon_{\mu~\mathrm{PID}}^{\rm data}$, $\varepsilon_{\mu~\mathrm{PID}}^{\rm MC}$, and $f_{\mu~\mathrm{PID}}^{\rm cor}$ obtained from samples I and II.}
	\label{tab:mupid}
	\begin{tabular}{lccc}\hline
		Samples & $\varepsilon_{\mu~\mathrm{PID}}^{\rm data}$ (\%) & $\varepsilon_{\mu~\mathrm{PID}}^{\rm MC}$ (\%) & $f_{\mu~\mathrm{PID}}^{\rm cor}$ \\\hline
		I	& $76.64\pm0.68$ & $81.04\pm0.21$ & $0.946\pm0.009$ \\
		II	& $76.85\pm0.30$ & $81.66\pm0.11$ & $0.941\pm0.004$ \\\hline
	\end{tabular}
\end{table*}


\begin{thebibliography}{**}
\bibitem{conjugate} Charge-conjugate modes are always implied unless stated explicitly.
\bibitem{muid} The $|\cos\theta_{\mu,\,i}|$
is equally divided as
$[0.2\times (i-1),\,0.2\times i)~(i=1,\, 2, \, 3,\, 4,\, {\rm or}~5)$.
In the first three $|\cos\theta_{\mu,\,i}|$ bins, 
we require $d_\mu$ greater than 17, 
$100\times p_\mu-(68+3\times i)$ and 33\,cm 
for the muons with $p_\mu \le 0.85+0.03\times i$,
$p_\mu\in(0.85+0.03\times i,\,1.01+0.03\times i)$ and $p_\mu\ge 1.01+0.03\times i$, respectively.
For other $|\cos\theta_{\mu,\,i}|$ bins, we require $d_\mu$ greater than 17\,cm uniformly.
\bibitem{dstauv} Without the $d_\mu$ requirement, there will be higher background from $D_s^+
\to\tau^+\nu_\tau, \tau^+\to\pi^+\bar\nu_\tau$.
\bibitem{bfweight} To be consistent with results at CLEO and BESIII, the BFs measured by Belle and BaBar have been further corrected by $f_{\rm cor}^{\rm rad}$.  The uncertainty in $f_{\rm cor}^{\rm rad}$ is
common for individual measurements, but will be taken into account in the averaged BF.
For the previous measurement in BESIII, we quote the BF without the LFU constraint.
\end{thebibliography}
\end{document}